\documentclass[11pt,preprint,nofootinbib,showpacs,preprintnumbers,amsmath,amssymb]{revtex4}

\usepackage{exscale}
\usepackage{relsize}
\usepackage{graphicx}
\usepackage{dcolumn}
\usepackage{bm}
\usepackage{multirow}
\usepackage{CJK}
\usepackage{diagbox}
\usepackage{subfigure}
\usepackage[T1]{fontenc}

\begin{document}

\begin{CJK*}{GBK}{}

\title{Phenomenological studies on the $\bar{B}^0\rightarrow [K^-\pi^+]_{S/V}[\pi^+\pi^-]_{V/S} \rightarrow K^-\pi^+\pi^+\pi^-$ decay}

\author{Jing-Juan Qi \footnote{e-mail: jjqi@mail.bnu.edu.cn}}
\affiliation{\small{Junior College, Zhejiang Wanli University, Zhejiang 315101, China}}

\author{Zhen-Yang Wang \footnote{Corresponding author, e-mail: wangzhenyang@nbu.edu.cn}}
\affiliation{\small{Physics Department, Ningbo University, Zhejiang 315211, China}}

\author{Zhu-Feng Zhang \footnote{e-mail: zhufengzhang@nbu.edu.cn}}
\affiliation{\small{Physics Department, Ningbo University, Zhejiang 315211, China}}

\author{Xin-Heng Guo \footnote{Corresponding author, e-mail: xhguo@bnu.edu.cn}}
\affiliation{\small{College of Nuclear Science and Technology, Beijing Normal University, Beijing 100875, China}}

\date{\today\\}

\begin{abstract}
Within the quasi-two-body decay model, we study the localized $CP$ violation and branching fraction of the four-body decay $\bar{B}^0\rightarrow [K^-\pi^+]_{S/V}[\pi^+\pi^-]_{V/S} \rightarrow K^-\pi^+\pi^-\pi^+$ when $K^-\pi^+$ and $\pi^-\pi^+$ pair invariant masses are $0.35<m_{K^-\pi^+}<2.04 \, \mathrm{GeV}$ and $0<m_{\pi^-\pi^+}<1.06\, \mathrm{GeV}$, with the pairs being dominated by the $\bar{K}^*_0(700)^0$, $\bar{K}^*(892)^0$, $\bar{K}^*(1410)^0$, $\bar{K}^*_0(1430)$ and $\bar{K}^*(1680)^0$, and $f_0(500)$, $\rho^0(770)$ , $\omega(782)$ and $f_0(980)$ resonances, respectively.  When dealing with the dynamical functions of these resonances, $f_0(500)$, $\rho^0(770)$, $f_0(980)$ and $\bar{K}^*_0(1430)$ are modeled with the Bugg model, Gounaris-Sakurai function, Flatt$\acute{\mathrm{e}}$ formalism and LASS lineshape, respectively, while others are described by the relativistic Breit-Wigner function. Adopting the end point divergence parameters $\rho_A\in[0,0.5]$ and $\phi_A\in[0,2\pi]$, our predicted results are $\mathcal{A_{CP}}(\bar{B}^0\rightarrow K^-\pi^+\pi^+\pi^-)\in[-0.383,0.421]$ and $\mathcal{B}(\bar{B}^0\rightarrow K^-\pi^+\pi^+\pi^-)\in[7.36,199.69]\times10^{-8}$ based on the hypothetical $q\bar{q}$ structures for the scalar mesons in the QCD factorization approach. Meanwhile, we calculate the $CP$ violating asymmetries and branching fractions of the two-body decays $\bar{B}^0\rightarrow SV(VS)$ and all the  individual four-body decays $\bar{B}^0\rightarrow SV(VS) \rightarrow K^-\pi^+\pi^-\pi^+$, respectively.  Our theoretical results for the two-body decays $\bar{B}^0\rightarrow \bar{K}^*(892)^0$$f_0(980)$, $\bar{B}^0\rightarrow \bar{K}^*_0(1430)^0$$\omega(782)$, $\bar{B}^0\rightarrow \bar{K}^*(892)^0f_0(980)$, $\bar{B}^0\rightarrow\bar{K}^*_0(1430)^0\rho$, and $\bar{B}^0\rightarrow\bar{K}^*_0(1430)^0\omega$ are consistent with the available experimental data, with the remaining predictions awaiting future examinations with the high precision. If they are confirmed by experiments in the future, the viewpoint that the scalar mesons have the $q\bar{q}$ composition should be well supported.
\end{abstract}


\maketitle
\end{CJK*}

\section{Introduction}
Differences in the behaviour of matter and antimatter ($CP$ violation) have been observed in several processes and, in particular, in charmless $B$ decays. The current understanding of the composition of matter in the Universe indicates that other mechanisms, beyond that proposed within the Standard Model (SM) of particle physics, could exist in order to account for the observed imbalance between the matter and antimatter. The study of $CP$ violating processes may therefore be used to test the corresponding SM predictions and place constraints on extensions of this framework. $CP$ violation is related to the weak complex phase in the Cabibbo-Kobayashi-Maskawa (CKM) matrix, which describes the mixing of different generations of quarks \cite{Cabibbo:1963yz, Kobayashi:1973fv}. Besides the weak phase, a large strong phase is also needed for direct $CP$ violation in decay processes. Usually, this large phase is provided by the short-distance or long-distance QCD interactions. The short-distance interactions are caused by QCD loop corrections, and the long-distance interaction which is more sensitive to the structure of the final states can be obtained through some phenomenological mechanisms.

Four-body decays of heavy mesons are hard to be investigated because of their complicated phase spaces and relatively smaller branching fractions. This leads to much less research in four-body dacays than that in two- or three- body decays\cite{Cheng:2020hyj,Aubert:2008bj,Garmash:2005rv,Aaij:2013sfa,Beneke:2003zv,Cheng:2013dua,Lu:2000em,Xiao:2011tx,Li:2015zra,Chang:2014rla,Zhang:2013oqa}. We have discussed localized CP violation and the branching fraction of the four-body decays $\bar{B}^0\rightarrow K^-\pi^+\pi^-\pi^+$ in Ref. \cite{Qi:2019nmn}. We have focused on the $\pi\pi$  and $K\pi$ invariant masses are near the masses of $f_0(500)$ and $\rho^0(770)$ mesons. Here we will further expand the area of our research to predict the study the CP violation and the branching fraction in the $\bar{B}^0$ four-body decays, which include more contributions from more different resonances. Specifically, the invariant mass of the $K^-\pi^+$ pair lies in the range $0.35<m_{K^-\pi^+}<2.04 \, \mathrm{GeV}$ which is dominated by the $\bar{K}^*_0(700)^0$, $\bar{K}^*(892)^0$, $\bar{K}^*(1410)^0$, $\bar{K}^*_0(1430)$ and $\bar{K}^*(1680)^0$ resonances, and that of the $\pi^-\pi^+$ pair is in the range $0<m_{\pi^-\pi^+}<1.06\, \mathrm{GeV}$ which includes the $f_0(500)$, $\rho^0(770)$ , $\omega(782)$ and $f_0(980)$ resonances. Meanwhile, studying the multibody decays can provide rich information for their intermediate resonances especially for the unclear compositions of scalar mesons. It is known that the identification of scalar mesons is difficult experimentally and the underlying structures of scalar mesons are not well established \cite{Godfrey:1998pd}. The investigation of their structures can improve our understanding about QCD and the quark confinement mechanism. The first charmless $B$ decay into a scalar meson that has been observed is $B\rightarrow f_0(980)K$. It was first measured by Belle in the charged $B$ decays to $K^\pm\pi^+\pi^\pm$ and a large branching fraction for the $f_0(980)K^\pm$ final states was found \cite{Abe:2002av} (updated in \cite{Garmash:2004wa}), which was subsequently confirmed by BABAR \cite{Aubert:2003mi}.  Studies of the mass spectra of scalar mesons as well as their strong and electromagnetic decays suggest that there exist two typical scenarios for their structures \cite{Cheng:2005nb, Cheng:2013fba}. In Scenario 1 (S1), the light scalar mesons with their masses below or near 1 GeV are treated as the lowest-lying $q\bar{q}$ states forming an SU(3) flavor nonet, including $f_0(500)$, $f_0(980)$, $\bar{K}^*_0(700)^0$ and $a_0(980)$, and those with masses near 1.5 $\mathrm{GeV}$ are suggested as the first corresponding excited states forming another SU(3) flavor nonet, such as $a_0(1450)$, $K^*_0(1430)$, $f_0(1370)$ and $f_0(1500)$ \cite{Jaffe:1976ig, Alford:2000mm}. In Scenario 2 (S2), the heavier nonet mesons are regarded as the ground states of $q\bar{q}$, while those lighter nonet ones are not regular mesons and might be four-quark states.

In 2019, LHCb collaboration study the $B^0\rightarrow \rho(770)^0K^*(892)^0$ decay within an quasi-two-body decay mode $B^0\rightarrow (\pi^+\pi^-)(K^+\pi^-)$ \cite{Aaij:2018bjo}.  In our work, we will adopt this mechanism to study the four-body decay $\bar{B}^0\rightarrow K^-\pi^+\pi^-\pi^+$, i.e. $\bar{B}^0\rightarrow \bar{\kappa}\rho\rightarrow K^-\pi^+\pi^-\pi^+$, $\bar{B}^0\rightarrow \bar{\kappa}\omega\rightarrow K^-\pi^+\pi^-\pi^+$, $\bar{B}^0\rightarrow\bar{K}^*(892)^0\sigma\rightarrow K^-\pi^+\pi^-\pi^+$, $\bar{B}^0\rightarrow \bar{K}^*(892)^0f_0(980)\rightarrow K^-\pi^+\pi^-\pi^+$, $\bar{B}^0\rightarrow \bar{B}^0\rightarrow\bar{K}^*(1410)^0\sigma\rightarrow K^-\pi^+\pi^-\pi^+$, $\bar{B}^0\rightarrow \bar{K}^*(1410)^0f_0(980)\rightarrow K^-\pi^+\pi^-\pi^+$, $\bar{B}^0\rightarrow \bar{K}^*_0(1430)^0\rho\rightarrow K^-\pi^+\pi^-\pi^+$, $\bar{B}^0\rightarrow \bar{K}^*_0(1430)^0\omega\rightarrow K^-\pi^+\pi^-\pi^+$, $\bar{B}^0\rightarrow \bar{K}^*(1680)^0\sigma\rightarrow K^-\pi^+\pi^-\pi^+$ and $\bar{B}^0\rightarrow \bar{K}^*(1680)^0f_0(980)\rightarrow K^-\pi^+\pi^-\pi^+$,  where the scalar mesons will be treated in S1 as mentioned above. We can then calculate the localized $CP$ violating asymmetries and branching fractions of the four-body decay $\bar{B}^0\rightarrow K^-\pi^+\pi^-\pi^+$. Besides we can also calculate the $CP$ asymmetries and branching fractions of the two-body decays $\bar{B}^0\rightarrow SV(VS)$ and all the individual four-body decays $\bar{B}^0\rightarrow SV(VS) \rightarrow K^-\pi^+\pi^-\pi^+$, respectively. In fact, with the great development of the large hadron collider beauty (LHCb) and Belle-II experiments, more and more decay modes involving one or two scalar states in the $B$ and $D$ meson decays are expected to be measured with the high precision in the future.

The remainder of this paper is organized as follows. Our theoretical framework are presented in Sect. ${\mathrm{\uppercase\expandafter{\romannumeral2}}}$. In Sect. ${\mathrm{\uppercase\expandafter{\romannumeral3}}}$, we give our numerical results. And we summarize our work in Sect ${\mathrm{\uppercase\expandafter{\romannumeral4}}}$. Appendix A collects the explicit formulas for all four-body decay amplitudes. The dynamical functions for the corresponding resonances are summarized in Appendix B. We also consider the $f_0(500)-f_0(980)$ mixing in Appendix C. Related theoretical parameters are listed in Appendix D.

\section{THEORETICAL FRAMEWORK}
\subsection{B decay in QCD factorization}
With the operator product expansion, the effective weak Hamiltonian for $B$ meson decays can be written as \cite{Beneke:2003zv}
\begin{equation}\label{Hamiltonian}
\mathcal{H}_{eff}=\frac{G_F}{\sqrt{2}}\sum_{p=u,c}\sum_{D=d,s}\lambda_{p}^{(D)}{(C_1Q_1^p+C_2Q_2^p+\sum_{i=3}^{10}C_iQ_i+C_{7\gamma}Q_{7\gamma}+C_{8g}Q_{8g})+h.c.},
\end{equation}
where $\lambda_p^{(D)}=V_{pb}V_{pD}^*$, $V_{pb}$ and $V_{pD}$ are the CKM matrix elements, $G_F$ represents the Fermi constant, $C_i$ $(i=1,2,\cdots,10)$ are the Wilson coefficients, $Q_{1,2}^p$ are the tree level operators and $Q_{3-10}$ are the penguin ones, and $Q_{7\gamma}$ and $Q_{8g}$ are the electromagnetic and chromomagnetic dipole operators, respectively. The explicit forms of the operators $Q_i$ are \cite{Beneke:2001ev}
\begin{equation}\label{qi}
\begin{split}
&Q_1^p=\bar{p}\gamma_\mu(1-\gamma_5)b\bar{D}\gamma^\mu(1-\gamma_5)p, \quad\quad\quad\quad\quad Q_2^p=\bar{p}_\alpha\gamma_\mu(1-\gamma_5)b_\beta\bar{D}_\beta\gamma^\mu(1-\gamma_5)p_\alpha, \\
&Q_3=\bar{D}\gamma_\mu(1-\gamma_5)b\sum_{q'}\bar{q'}\gamma^\mu(1-\gamma_5)q',\quad\quad\quad Q_4=\bar{D}_\alpha\gamma_\mu(1-\gamma_5)b_\beta\sum_{q'}\bar{q'}_\beta\gamma^\mu(1-\gamma_5)q'_\alpha, \\
&Q_5=\bar{D}\gamma_\mu(1-\gamma_5)b\sum_{q'}\bar{q'}\gamma^\mu(1+\gamma_5)q',\quad\quad\quad Q_6=\bar{D}_\alpha\gamma_\mu(1-\gamma_5)b_\beta\sum_{q'}\bar{q'}_\beta\gamma^\mu(1+\gamma_5)q'_\alpha,\\
&Q_7=\frac{3}{2}\bar{D}\gamma_\mu(1-\gamma_5)b\sum_{q'}e_{q'}\bar{q'}\gamma^\mu(1+\gamma_5)q',\quad\, Q_8=\frac{3}{2}\bar{D}_\alpha\gamma_\mu(1-\gamma_5)b_\beta\sum_{q'}e_{q'}\bar{q'}_\beta\gamma^\mu(1+\gamma_5)q'_\alpha,\\
&Q_9=\frac{3}{2}\bar{D}\gamma_\mu(1-\gamma_5)b\sum_{q'}e_{q'}\bar{q'}\gamma^\mu(1-\gamma_5)q',\quad\, Q_{10}=\frac{3}{2}\bar{D}_\alpha\gamma_\mu(1-\gamma_5)b_\beta\sum_{q'}e_{q'}\bar{q'}_\beta\gamma^\mu(1-\gamma_5)q'_\alpha,\\
&Q_{7\gamma}=\frac{-e}{8\pi^2}m_b\bar{s}\sigma_{\mu\nu}(1+\gamma_5)F^{\mu\nu}b,\quad \quad\quad\quad\quad\quad Q_{8g}=\frac{-g_s}{8\pi^2}m_b\bar{s}\sigma_{\mu\nu}(1+\gamma_5)G^{\mu\nu}b,\\
\end{split}
\end{equation}
where $\alpha$ and $\beta$ are color indices, $q'=u,d,s,c$ or $b$ quarks.

Within the framework of QCD factorization \cite{Beneke:2003zv,Beneke:2001ev}, the effective Hamiltonian matrix elements are written in the form
\begin{equation}
\langle{M_1M_2}|\mathcal{H}_{eff}|B\rangle=\sum_{p=u,c}\lambda_{p}^{(D)}\langle{M_1M_2}|\mathcal{T}_A^p+\mathcal{T}_B^p|B\rangle,
\end{equation}
where $\mathcal{T}_A^p$ describes the contribution from naive factorization, vertex correction, penguin amplitude and spectator scattering expressed in terms of the parameters $a_i^p$, while $\mathcal{T}_B^p$ contains annihilation topology amplitudes characterized by the annihilation parameters $b_i^p$.

The flavor parameters $a_i^p$ are basically the Wilson coefficients in conjunction with short-distance nonfactorizable corrections such as vertex corrections and hard spectator interactions. In general, they have the expressions \cite{Beneke:2003zv}
\begin{equation}\label{a}
\begin{split}
a_i^p{(M_1M_2)}&={(c'_i+\frac{c'_{i\pm1}}{N_c})}N_i{(M_2)}+\frac{c'_{i\pm1}}{N_c}\frac{C_F\alpha_s}{4\pi}{\bigg[V_i{(M_2)}+\frac{4\pi^2}{N_c}H_i{(M_1M_2)}\bigg]+P_i^p{(M_2)}},
\end{split}
\end{equation}
where $c'_i$ are effective Wilson coefficients which are defined as $c_i(m_b)\langle O_i(m_b)\rangle=c'_i\langle O_i\rangle^{tree}$, with $\langle O_i\rangle^{tree}$ being the matrix element at the tree level, the upper (lower) signs apply when $i$ is odd (even), $N_i{(M_2)}$ is leading-order coefficient, $C_F={(N_c^2-1)}/{2N_c}$ with $N_c=3$, the quantities $V_i{(M_2)}$ account for one-loop vertex corrections, $H_i{(M_1M_2)}$ describe hard spectator interactions with a hard gluon exchange between the emitted meson and the spectator quark of the $B$ meson, and $P_i^p{(M_1M_2)}$ are from penguin contractions \cite{Beneke:2003zv}.

The weak annihilation contributions to $B\rightarrow M_1M_2$ can be described in terms of $b_i$ and $b_{i,EW}$, which have the following expressions:
\begin{equation}\label{b}
\begin{split}
&b_1=\frac{C_F}{N_c^2}c'_1A_1^i, \quad b_2=\frac{C_F}{N_c^2}c'_2A_1^i, \\
&b_3^p=\frac{C_F}{N_c^2}\bigg[c'_3A_1^i+c'_5(A_3^i+A_3^f)+N_cc'_6A_3^f \bigg],\quad b_4^p=\frac{C_F}{N_c^2}\bigg[c'_4A_1^i+c'_6A_2^i \bigg], \\
&b_{3,EW}^p=\frac{C_F}{N_c^2}\bigg[c'_9A_1^i+C'_7(A_3^i+A_3^f)+N_cc'_8A_3^f \bigg],\\
&b_{4,EW}^p=\frac{C_F}{N_c^2}\bigg[c'_{10}A_1^i+c'_8A_2^i \bigg],
\end{split}
\end{equation}
where the subscripts 1, 2, 3 of $A_n^{i,f}(n=1,2,3)$ stand for the annihilation amplitudes induced from $(V-A)(V-A)$, $(V-A)(V+A)$, and $(S-P)(S+P)$ operators, respectively, the superscripts $i$ and $f$ refer to gluon emission from the initial- and final-state quarks, respectively. The explicit
expressions for $A_n^{i,f}$ can be found in Ref. \cite{Cheng:2007st}.

In the expressions for the spectator and annihilation corrections, there are end-point divergences $X=\int_0^1 dx/(1-x)$. The QCD factorization
approach suffers from these end-point divergences, which can be parametrized as \cite{Cheng:2005nb}
\begin{equation}\label{XA}
X_{A,H}=(1+\rho_{A,H} e^{i\phi_{A,H}})\ln\frac{m_B}{\Lambda_h},
\end{equation}
with $\Lambda_h$ being a typical scale of order 500 $\mathrm{MeV}$, $\rho_{A,H}$ an unknown real parameter and $\phi_{A,H}$ the free strong phase in the range $[0,2\pi]$.

\subsection{Four-body decay amplitudes}
For the four-body decay $\bar{B}^0\rightarrow K^-\pi^+\pi^-\pi^+$, we consider the two-body cascade decays mode $\bar{B}^0\rightarrow[K^-\pi^+]_{S/V}[\pi^-\pi^+]_{V/S}\rightarrow K^-\pi^+\pi^-\pi^+$. Within the QCDF framework in Ref. \cite{Beneke:2003zv}, we can deduce the two-body weak decay amplitudes of $\bar{B}^0\rightarrow [K^-\pi^+]_{S/V}[\pi^-\pi^+]_{V/S}$, which are

\begin{eqnarray}\label{amplitude131}
\begin{split}
\mathcal{M}(\bar{B}^0\rightarrow \bar{K}^{*0}_{0i}\rho
)&=iG_F\sum_{p=u,c}\lambda_p^{(s)}\bigg\{\bigg[\delta_{pu}\alpha_2(\bar{K}^{*0}_{0i}\rho)
+\frac{3}{2}\alpha_{3,EW}^p(\bar{K}^{*0}_{0i}\rho)\bigg]
f_\rho m_\rho \varepsilon_\rho^*\cdot p_BF_1^{\bar{B}^0 \bar{K}^{*0}_{0i}}(m_{\rho}^2)\\
&+\bigg[\alpha_4^p(\rho\bar{K}^{*0}_{0i})-\frac{1}{2}\alpha_{4,EW}^p(\rho\bar{K}^{*0}_{0i})\bigg]
\bar{f}_{\bar{K}^{*0}_{0i}}m_\rho \varepsilon_\rho^*\cdot p_BA_0^{\bar{B}^0\rho}(m_{\bar{K}^{*0}_{0i}}^2)\\
&+\bigg[\frac{1}{2}b_3^p(\rho\bar{K}^{*0}_{0i})-\frac{1}{4}b_{3,EW}^p(\rho\bar{K}^{*0}_{0i})\bigg]
f_{\bar{B}^0}f_\rho \bar{f}_{\bar{K}^{*0}_{0i}}\bigg\},\\
\end{split}
\end{eqnarray}

\begin{eqnarray}\label{amplitude141}
\begin{split}
\mathcal{M}(\bar{B}^0\rightarrow \bar{K}^{*0}_{0i}\omega)&=iG_F\sum_{p=u,c}\lambda_p^{(s)}\bigg\{\bigg[\delta_{pu}\alpha_2(\bar{K}^{*0}_{0i}\omega)+2\alpha_3^p(\bar{K}^{*0}_{0i}\omega)
+\frac{1}{2}\alpha_{3,EW}^p(\bar{K}^{*0}_{0i}\omega)\bigg]
f_\omega m_\omega \varepsilon_\omega^*\cdot p_BF_1^{\bar{B}^0 \bar{K}^{*0}_{0i}}(m_{\omega}^2)\\
&+\bigg[\frac{1}{2}\alpha_{4,EW}^p(\omega\bar{K}^{*0}_{0i})-\alpha_4^p(\omega\bar{K}^{*0}_{0i})\bigg]
\bar{f}_{\bar{K}^{*0}_{0i}}m_\omega \varepsilon_\omega^*\cdot p_BA_0^{\bar{B}^0\omega}(m_{\bar{K}^{*0}_{0i}}^2)\\
&+\bigg[\frac{1}{4}b_{3,EW}^p(\omega\bar{K}^{*0}_{0i})-\frac{1}{2}b_3^p(\omega\bar{K}^{*0}_{0i})\bigg]
f_{\bar{B}^0}f_\rho \bar{f}_{\bar{K}^{*0}_{0i}}\bigg\},\\
\end{split}
\end{eqnarray}
with $\bar{K}^{*0}_{0i}=\bar{K}^*_0(700)^0,\bar{K}^*_0(1430)^0$ corresponding to $i=1,2$, respectively, and

\begin{eqnarray}\label{amplitude211}
\begin{split}
\mathcal{M}(\bar{B}^0\rightarrow \bar{K}^{*0}_if_{0j})&=-iG_F\sum_{p=u,c}\lambda_p^{(s)}\bigg\{\bigg[\delta_{pu}\alpha_2(\bar{K}^{*0}_if_{0j})+2\alpha_3^p(\bar{K}^{*0}_if_{0j})+\frac{1}{2}\alpha_{3,EW}^p(\bar{K}^{*0}_if_{0j})\bigg]\\
&\times\bar{f}_{f_{0j}^n}m_{\bar{K}^{*0}_i} \varepsilon_{\bar{K}^{*0}_i}^*\cdot p_BA_0^{\bar{B}^0 \bar{K}^{*0}_i}(m_{f_{0j}}^2)+\bigg[\sqrt{2}\alpha_3^p(\bar{K}^{*0}_if_{0j})+\sqrt{2}\alpha_4^p(\bar{K}^{*0}_if_{0j})\\
&-\frac{1}{\sqrt{2}}\alpha_{3,EW}^p(\bar{K}^{*0}_if_{0j})-\frac{1}{\sqrt{2}}\alpha_{4,EW}^p(\bar{K}^{*0}_if_{0j})\bigg]\bar{f}_{f_{0j}^s}m_{\bar{K}^{*0}_i} \varepsilon_{\bar{K}^{*0}_i}^*\cdot p_BA_0^{\bar{B}^0\bar{K}^{*0}_i}(m_{f_{0j}}^2)\\
&+\bigg[\frac{1}{2}\alpha_{4,EW}^p(f_{0j}\bar{K}^{*0}_i)-\alpha_4^p(f_{0j}\bar{K}^{*0}_i)\bigg]f_{\bar{K}^{*0}_i}m_{\bar{K}^{*0}_i} \varepsilon_{\bar{K}^{*0}_i}^*\cdot p_B
F_1^{\bar{B}^0f_{0j}}(m_{\bar{K}^{*0}_i}^2)+\bigg[\frac{1}{\sqrt{2}}b_3^p(\bar{K}^{*0}_if_{0j})\\
&-\frac{1}{2\sqrt{2}}b_{3,EW}^p(\bar{K}^{*0}_if_{0j})\bigg]f_{\bar{B}^0}f_{\bar{K}^{*0}_i}\bar{f}_{f_{0j}}^s+\bigg[\frac{1}{2}b_3^p(f_{0j}\bar{K}^{*0}_i)-\frac{1}{4}b_{3,EW}^p(f_{0j}\bar{K}^{*0}_i)\bigg]f_{\bar{B}^0}f_{\bar{K}^{*0}_i}\bar{f}_{f_{0j}}^n\bigg\},\\
\end{split}
\end{eqnarray}
with $\bar{K}^{*0}_i=\bar{K}^*(892)^0,\bar{K}^*(1410)^0,\bar{K}^*(1680)^0$ corresponding to $i=1,2,3$, respectively, $f_{0j}=f_0(500)$, $f_0(980)$ when $j=1,2$, respectively. In Eqs. (\ref{amplitude131})-(\ref{amplitude211}), $F_1^{\bar{B}^0 \rightarrow S}(m_V^2)$ and $A_0^{\bar{B}^0\rightarrow V}(m_S^2)$ are the form factors for $\bar{B}^0$ to scalar and vector meson transitions, respectively, $f_V$,  $\bar{f}_S$, and $f_{\bar{B}^0}$ are decay constants of vector, scalar, and $\bar{B}^0$ mesons, respectively, $\bar{f}_{f_{0j}}^s$ and $\bar{f}_{f_{0j}}^n$ are decay constants of $f_{0j}$ mesons coming from the up and strange quark components, respectively.

In the framework of the two two-body decays, the four-body decay can be factorized into three pieces as the following:
\begin{equation} \label{HKrho1}
\begin{split}
\mathcal{M}(\bar{B}^0\rightarrow [K^-\pi^+]_S[\pi^-\pi^+]_V\rightarrow K^-\pi^+\pi^-\pi^+)=\frac{\langle SV|\mathcal{H}_{eff}|\bar{B}^0\rangle \langle K^-\pi^+|\mathcal{H}_{S K^-\pi^+}|S\rangle \langle \pi^-\pi^+|\mathcal{H}_{V \pi^-\pi^+}|V\rangle}{s_{S}s_{V}},
\end{split}
\end{equation}
and
\begin{equation} \label{HKrho2}
\begin{split}
\mathcal{M}(\bar{B}^0\rightarrow [K^-\pi^+]_{V}[\pi^-\pi^+]_{S}\rightarrow K^-\pi^+\pi^-\pi^+)=\frac{\langle VS|\mathcal{H}_{eff}|\bar{B}^0\rangle \langle K^-\pi^+|\mathcal{H}_{V K^-\pi^+}|V\rangle \langle \pi^-\pi^+|\mathcal{H}_{S\pi^-\pi^+}|S\rangle}{s_Vs_S},
\end{split}
\end{equation}
where $\mathcal{H}_{eff}$ is the effective weak Hamiltonian, $\langle M_1M_2|\mathcal{H}_s|V\rangle=g_{VM_1M_2}(p_{M_1}-p_{M_2})\cdot\epsilon_V$ and $\langle M_1M_2|\mathcal{H}_s|S\rangle=g_{SM_1M_2}$, $g_{VM_1M_2}$ and $g_{SM_1M_2}$ are the strong coupling constants of the corresponding vector and scalar mesons decays, $s_{S/V}$ are the reciprocal of dynamical functions $T_{S/V}$ for the corresponding resonances. The specific kinds and expressions of $T_{S/V}$ are given in the fifth column of Table \ref{masses widths} and Appendix C, respectively.

When considering the contributions from $\bar{B}^0\rightarrow [K^-\pi^+]_S[\pi^-\pi^+]_V\rightarrow K^-\pi^+\pi^-\pi^+ $ and $\bar{B}^0\rightarrow [K^-\pi^+]_{V}[\pi^-\pi^+]_{S}\rightarrow K^-\pi^+\pi^-\pi^+$ channels as listed in Eqs. (\ref{HKrho1}) and (\ref{HKrho2}), the total decay amplitude of the $\bar{B}^0\rightarrow K^-\pi^+\pi^+\pi^-$ decay can be written as \footnote {There could be a relative strong phase $\delta$ between the two interference amplitudes, which values depend on experimental data and theoretical models. Since little information about $\delta$ can be provided by experiments, we choose to adopt the same method as that in Refs. \cite{Cheng:2013dua,Cheng:2014uga,Li:2014fla}, i.e. setting $\delta=0$.}
\begin{equation}\label{A}
\mathcal{M}=\mathcal{M}(\bar{B}^0\rightarrow [K^-\pi^+]_S[\pi^-\pi^+]_V\rightarrow K^-\pi^+\pi^-\pi^+ )+\mathcal{M}(\bar{B}^0\rightarrow [K^-\pi^+]_{V}[\pi^-\pi^+]_{S}\rightarrow K^-\pi^+\pi^-\pi^+).
\end{equation}

\subsection{Kinematics of the four-body decay and localized CP violation}
One can use the five variables to describe the four-body decay kinematics \cite{Bijnens:1994ie,Kang:2013jaa}, which are the invariant mass squared of the $K\pi$ system $s_{K\pi}=(p_1+p_2)^2=m_{K\pi}^2$, the invariant mass squared of the $\pi\pi$ system $s_{\pi\pi}=(p_3+p_4)^2=m_{\pi\pi}^2$, the angles $\theta_\pi$, $\theta_K$ and $\phi$, where $\theta_\pi$ is the angle of the $\pi^+$ in the $\pi^-\pi^+$ center-of-mass frame $\Sigma_{\pi\pi}$ with respect to the pions' line of flight in the $\bar{B}^0$ rest frame $\Sigma_{\bar{B}^0}$, $\theta_K$ is the angle of the $K^-$ in the $K\pi$ center-of-mass system $\Sigma_{K\pi}$ with respect to the $K\pi$ line of flight in $\Sigma_{\bar{B}^0}$ and $\phi$ is the angle between the $K\pi$ and $\pi\pi$ planes, respectively. Their physical ranges are

\begin{equation}\label{Mand}
\begin{split}
4m_{\pi\pi}^2&\leq s_{\pi\pi}\leq(m_{\bar{B}^0}-m_{K\pi})^2,\\
(m_K+m_\pi)^2&\leq s_{K\pi}\leq(m_{\bar{B}^0}-\sqrt{s_{\pi\pi}})^2,\\
0&\leq\theta_\pi,\theta_K\leq\pi,\quad 0\leq\phi\leq2\pi.\\
\end{split}
\end{equation}

Instead of the individual momenta $p_1$, $p_2$, $p_3$, $p_4$, it is more convenient to use the following kinematic variables
\begin{equation}\label{kinvar}
\begin{split}
 P&=p_1+p_2,\quad Q=p_1-p_2,\\
 L&=p_3+p_4,\quad N=p_3-p_4.\\
 \end{split}
 \end{equation}
It follows that
\begin{equation}\label{kinvar}
\begin{split}
 P^2&=s_{K\pi},\quad Q^2=2(p_K^2+p_\pi^2)-s_{K\pi},\quad L^2=s_{\pi\pi},\\
 P\cdot L&=\frac{1}{2}(m_{\bar{B}^0}^2-s_{K\pi}-s_{\pi\pi}),\quad P\cdot N=X\cos\theta_\pi,\quad L\cdot Q=\sigma(s_{K\pi})X\cos\theta_K,\\
 \end{split}
 \end{equation}
where
\begin{equation}\label{sigmaK}
\begin{split}
 \sigma(s_{K\pi})=\sqrt{1-(m_K^2+m_\pi^2)/s_{K\pi}},\\
 \end{split}
 \end{equation}
and the function $X$ is defined as
\begin{equation}\label{X}
\begin{split}
 X(s_{K\pi},s_{\pi\pi})&=\bigg[(P\cdot L)^2-s_{K\pi}s_{\pi\pi}\bigg]^{1/2}=\frac{1}{2}\lambda^{1/2}(m_{\bar{B}^0}^2,s_{K\pi},s_{\pi\pi}),\\
\lambda(x,y,z)&=(x-y-z)^2-4yz.\\
 \end{split}
 \end{equation}

With the decay amplitude, one can get the decay rate of the four-body decay \cite{Cheng:2017qpv},
\begin{equation} \label{B}
d^5\Gamma=\frac{1}{4(4\pi)^6m_{\bar{B}^0}^3}\sigma(s_{\pi\pi})X(s_{\pi\pi},s_{K\pi})\sum_{\mathrm{spins}}|\mathcal{M}|^2d\Omega,
\end{equation}
where $\sigma(s_{\pi\pi})=\sqrt{1-4m_\pi^2/s_{\pi\pi}}$, $\Omega$ represents the phase space with $d\Omega=ds_{\pi\pi}ds_{K\pi}dcos\theta_\pi dcos\theta_Kd\phi$.

The differential $CP$ asymmetry parameter and the localized integrated $CP$ asymmetry take the following forms
 \begin{equation}\label{CP asymmetry parameter}
\mathcal{A_{CP}}=\frac{|\mathcal{M}|^2-|\bar{\mathcal{M}}|^2}{|\mathcal{M}|^2+|\bar{\mathcal{M}}|^2},
 \end{equation}
and
  \begin{equation}\label{localized CP}
\mathcal{A^\mathrm{\Omega}_{CP}}=\frac{\int d\Omega(|\mathcal{M}|^2-|\bar{\mathcal{M}}|^2)}{\int d\Omega(|\mathcal{M}|^2+|\bar{\mathcal{M}}|^2)},
 \end{equation}
respectively.

\section{NUMERICAL RESULTS}
Using the large energy effective theory (LEET) techniques, Refs. \cite{Hatanaka:2009gb,Hatanaka:2009sj} formulate the $\bar{B}^0\rightarrow \bar{K}^{*0}_J$ ($J\geq1$) form factors in the large recoil region. All the form factors can be expressed in terms of two independent LEET functions, $\xi_\perp$ and $\xi_\parallel$. Explicitly, we have
\begin{equation}
A_0^{\bar{B}^0\rightarrow \bar{K}^{*0}_J}(q^2)\bigg(\frac{|\vec{p}_{\bar{K}^{*0}_J}|}{m_{\bar{K}^{*0}_J}}\bigg)^{J-1}\simeq\bigg(1-\frac{m^2_{\bar{K}^{*0}_J}}{m_{\bar{B}^0}E}\bigg)\xi^{\bar{K}^{*0}_J}_\parallel(q^2)+\frac{m_{\bar{K}^{*0}_J}}{m_{\bar{B}^0}}\xi_\perp^{\bar{K}^{*0}_J}(q^2),
\end{equation}
where we have used $|\vec{p}_{\bar{K}^{*0}_J}|/E\simeq1$, $|\vec{p}_{\bar{K}^{*0}_J}|$ is the magnitude of the
three momentum of $\bar{K}^{*0}_J$ in the rest frame of the $\bar{B}^0$ meson. With $\xi^{\bar{K}^*(1410)^0}_\parallel(0)=0.22\pm0.03$, $\xi^{\bar{K}^*(1410)^0}_\perp(0)=0.28\pm0.04$, $\xi^{\bar{K}^*(1680)^0}_\parallel(0)=0.18\pm0.03$ and $\xi^{\bar{K}^*(1680)^0}_\perp(0)=0.24\pm0.05$ derived from the Bauer-Stech-Wirbel (BSW) model \cite{Wirbel:1985ji}, we can obtain $A_0^{\bar{B}^0\rightarrow \bar{K}^*(1410)^0}(0)= 0.26\pm0.0275$ and $A_0^{\bar{B}^0\rightarrow \bar{K}^*(1680)^0}(0)=0.2154\pm0.0281$, respectively. In our work, all the form factors are evaluated at $q^2=0$ due to the smallness of $m_{V}^2$ and $m_{S}^2$ compared with $m_{\bar{B}^0}^2$. We also set $F^{\bar{B}^0\rightarrow\kappa}(0)=0.3$ and assign its uncertainty to be $\pm0.1$ for simplicity. As for the decay constants and Gegenbauer moments of the $\bar{K}^*(1410)^0$ and the $\bar{K}^*(1680)^0$ mesons, we assume they have the same central values as that of $\bar{K}^*(892)^0$ and assign their uncertainties to be $\pm0.1$ \cite{Qi:2018syl}.

As for the scalar meson, we adopt Scenario 1 in Ref. \cite{Cheng:2005nb}, in which those with masses below or near 1 GeV ($\sigma$, $f_0(980)$, $\kappa$) and near 1.5 $\mathrm{GeV}$ ($K^*_0(1430)$) are suggested as the lowest-lying $q\bar{q}$ states and the first excited state, respectively. When dealing with the decay constants of $f_{0j}$ mesons, we consider the $f_0(500)-f_0(980)$ mixing with the mixing angle $|\varphi_m|=17^0$ (see Appendix A for details). With the QCDF approach, we have obtained the amplitudes of the two-body decays $\bar{B}^0\rightarrow SV$ and $\bar{B}^0\rightarrow VS$, which are listed in Eqs. {(\ref{amplitude131})-(\ref{amplitude211})}. Generally, the end-point divergence parameter $\rho_A$ is constrained in the range of $[0,1]$ and $\phi_A$ is treated as a free strong phase. The experimental data of $B$ two-body decays can provide important information to restrict the ranges of these two parameters. In fact, compared with the $B\rightarrow PV/VP/PP$ decays, there are much less experimental data for the $B \rightarrow VS/PS$ and $B\rightarrow SV/SP$ decays, so the values of $\rho_A$ and $\phi_A$ have not been determined well in these decays. Thus we adopt $\rho_{A,H}<0.5$ and $0\leq\phi_{A,H}\leq 2\pi$ as in Refs. \cite{Cheng:2005nb,Cheng:2007st}. With more accumulation of experimental data, both of them could be defined in small regions in the future.
\begin{table}[tb]
\renewcommand{\arraystretch}{1.2}
\centering
\caption{Direct $CP$ asymmetries (in units of $10^{-2}$) of the two-body decays $\bar{B}^0\rightarrow [K^-\pi^+]_{S/V}[\pi^+\pi^-]_{V/S}$. The experimental branching fractions are taken from Ref. \cite{Amhis:2014hma}. The theoretical errors come from the uncertainties of the form factors, decay constants, Gegenbauer moments and divergence parameters.}
\begin{tabular*}{\textwidth}{@{\extracolsep{\fill}}ccccc}
\hline
\hline
Decay mode                     &BABAR                            &PDG \cite{Tanabashi:2018oca}     &\cite{Cheng:2013fba}           &This work\\
\hline
$\bar{\kappa}$$\rho$                 &$-$                              &$-$                             &$-$                                &$-12.65\pm3.20$\\
$\bar{\kappa}$$\omega$               &$-$                              &$-$                             &$-$                                &$15.19\pm6.92$\\
$\bar{K}^*(892)^0$$\sigma$      &$-$                              &$-$                             &$-$                                &$27.84\pm11.60$\\
$\bar{K}^*(892)^0$$f_0(980)$   &$7\pm10\pm2$                     &$7\pm10$                         &$-$                                &$8.52\pm1.27$\\
$\bar{K}^*(1410)^0$$\sigma$    &$-$                              &$-$                             &$-$                                &$0.27\pm0.11$\\
$\bar{K}^*(1410)^0$$f_0(980)$   &$-$                              &$-$                             &$-$                                &$-1.76\pm0.23$\\
$\bar{K}^*_0(1430)^0$$\rho$    &$-$                              &$-$                    &$0.54^{+0.45+0.02+3.76}_{-0.46-0.02-1.80}$   &$5.98\pm1.33$\\
$\bar{K}^*_0(1430)^0$$\omega$  &$-7\pm9\pm2$                     &$-$             &$0.03^{+0.37+0.01+0.29}_{-0.35-0.01-3.00}$          &$-10.48\pm4.19$\\
$\bar{K}^*(1680)^0$$\sigma$    &$-$                              &$-$                        &$-$                                   &$2.83\pm0.81$\\
$\bar{K}^*(1680)^0$$f_0(980)$  &$-$                              &$-$                       &$-$                                    &$-2.98\pm0.26$\\
\hline
\hline
\end{tabular*}\label{cpviolation1}
\end{table}

Substituting Eqs. (\ref{amplitude131})- (\ref{amplitude211}) into Eq. (\ref{CP asymmetry parameter}), we obtain the $CP$ violating asymmetries of the two-body decays $\bar{B}^0\rightarrow SV$ and $\bar{B}^0\rightarrow VS$ with the parameters given in Table \ref{masses widths} and Appendix F, which are listed in Table \ref{cpviolation1}. From Table \ref{cpviolation1}, one can see our theoretical results for the $CP$ asymmetries of $\bar{B}^0\rightarrow \bar{K}^*(892)^0 f_0(980)$ and $\bar{B}^0\rightarrow \bar{K}^*_0(1430)^0$$\omega$ are consistent with the data from BABAR Collaboration. However, the predicted central values of the $CP$ asymmetries of the $\bar{B}^0\rightarrow \bar{K}^*_0(1430)^0 \rho$ and $\bar{B}^0\rightarrow \bar{K}^*_0(1430)^0\omega$ are larger than those in Ref. \cite{Cheng:2013fba}. The main difference between our work and Ref. \cite{Cheng:2013fba} is the structure of the $\bar{K}^*_0(1430)^0$ meson, which is explored in S1 in our work and S2 in Ref. \cite{Cheng:2013fba}, respectively. Besides, we predict the $CP$ asymmetries of some other channel decays. We find the signs of the $CP$ asymmetries are negative in $\bar{B}^0\rightarrow \bar{\kappa}\rho$, $\bar{B}^0\rightarrow \bar{K}^*(1410)^0 f_0(980)$ and $\bar{B}^0\rightarrow \bar{K}^*(1680)^0 f_0(980)$ decays, with the first one being one order larger than the other two. For the positive values of the $CP$ asymmetries in our work, those for the $\bar{B}^0\rightarrow\bar{\kappa}\omega$ and $\bar{B}^0\rightarrow\bar{K}^*(892)^0\sigma$ decays are also one order larger than the others. Moreover, we also calculate the branching fractions of the two-body decays $\bar{B}^0\rightarrow SV$ and $\bar{B}^0\rightarrow VS$ which are listed in Table \ref{brachingfractions1}. As can be seen from Table \ref{brachingfractions1}, our results are consistent with the available experimental data for the $\bar{B}^0\rightarrow \bar{K}^*(892)^0f_0(980)$, $\bar{B}^0\rightarrow\bar{K}^*_0(1430)^0\rho$ and $\bar{B}^0\rightarrow\bar{K}^*_0(1430)^0\omega$ decays. Meanwhile, we find the magnitudes of the branching fractions are of order $10^{-5}$ for $\bar{B}^0\rightarrow\bar{K}^*(892)^0f_0(980)$, $\bar{B}^0\rightarrow\bar{K}^*(1410)^0\sigma$ and $\bar{B}^0\rightarrow\bar{K}^*(1410)^0f_0(980)$, but of $10^{-6}$ for
 $\bar{B}^0\rightarrow\bar{\kappa}\rho$, $\bar{B}^0\rightarrow\bar{\kappa}\omega$, $\bar{B}^0\rightarrow\bar{K}^*_0(1430)^0\rho$ and $\bar{B}^0\rightarrow\bar{K}^*_0(1430)^0\omega$. We note that the predicted branching fraction of $\bar{B}^0\rightarrow\bar{K}^*(892)^0$$\sigma$ is the smallest one with the order of $10^{-7}$.
\begin{table}[tb]
\renewcommand{\arraystretch}{1.2}
\centering
\caption{Branching fractions (in units of $10^{-6}$) of the two-body decays $\bar{B}^0\rightarrow [K^-\pi^+]_{S/V}[\pi^+\pi^-]_{V/S}$. We have used $\mathcal{B}(f_0(980)\rightarrow \pi^+\pi^-)=0.5$ to obtain the experimental branching fractions for $f_0(980)V$. The theoretical errors come from the uncertainties of the form factors, decay constants Gegenbauer moments and divergence parameters.}
\begin{tabular*}{\textwidth}{@{\extracolsep{\fill}}cccccccc}
\hline
\hline
Decay mode    &BABAR  & Belle  & LHCb\cite{Cheng:2007st}  &PDG \cite{Tanabashi:2018oca}   &QCDF\cite{Cheng:2013fba} &pQCD\cite{Kim:2009dg,Zhang:2010kw}  &This work\\
\hline
$\bar{\kappa}$$\rho$        &$-$       &$-$          &$-$                 &$-$                            &$-$                      &$-$               &$1.27\pm0.53$\\
$\bar{\kappa}$$\omega$      &$-$       &$-$          &$-$                 &$-$                            &$-$                      &$-$               &$4.12\pm1.78$\\
$\bar{K}^*(892)^0$$\sigma$  &$-$       &$-$          &$-$                 &$-$                            &$-$                      &$-$               &$0.13\pm0.02$\\
$\bar{K}^*(892)^0$$f_0(980)$&$11.4\pm1.4$&$<4.4$&$-$&$7.8^{+4.2}_{-3.6}$&$9.1^{+1.0+1.0+5.3}_{-0.4-0.5-0.7}$ &$11.2\sim13.7$          &$10.81\pm3.39$\\
$\bar{K}^*(1410)^0$$\sigma$ &$-$       &$-$          &$-$                 &$-$                            &$-$                      &$-$               &$22.58\pm7.13$\\
$\bar{K}^*(1410)^0$$f_0(980)$&$-$       &$-$         &$-$                 &$-$                            &$-$                      &$-$               &$16.434\pm5.71$\\
$\bar{K}^*_0(1430)^0$$\rho$&$27\pm4\pm2\pm3$ &$-$&$10.0^{+2.4+0.5+12.1}_{-2.0-0.4-3.1}$&$27.0\pm6.0$ &$4.1^{+1.1+0.2+2.6}_{-1.0-0.2-0.1}$&$4.8^{+1.1+1.0+0.3}_{-0.0-1.0-0.3}$     &$8.96\pm1.75$\\
$\bar{K}^*_0(1430)^0$$\omega$&$6.4^{+1.4+0.3+4.0}_{-1.2-0.2-0.9}$&$-$&$-$&$16.0\pm3.4$&$9.3^{+2.7+0.3+3.9}_{-2.2-0.3-1.3}$&$9.3^{+2.1+3.6+1.2}_{-2.0-2.9-1.0}$&$4.43\pm1.69$\\
$\bar{K}^*(1680)^0$$\sigma$  &$-$       &$-$         &$-$                 &$-$                            &$-$                      &$-$               &$29.49\pm10.36$\\
$\bar{K}^*(1680)^0$$f_0(980)$&$-$       &$-$         &$-$                 &$-$                            &$-$                      &$-$               &$20.34\pm9.04$\\
\hline
\hline
\end{tabular*}\label{brachingfractions1}
\end{table}

As mentioned in the abstract, for different intermediate resonance states, we use different models to deal with their dynamical functions which are listed in Table \ref{masses widths} and Appendix D in detail, where $\sigma$, $\rho^0(770)$, $f_0(980)$ and $\bar{K}^*_0(1430)$ are modeled with the Bugg model \cite{Bugg:2006gc}, Gounaris-Sakurai function \cite{Gounaris:1968mw}, Flatt$\acute{\mathrm{e}}$ formalism \cite{Flatte:1976xv} and LASS lineshape \cite{Aston:1987ir,Aubert:2005ce,Aaij:2018bla}, respectively, while others are described by the relativistic Breit-Wigner function \cite{Chen:2020sog}. Inserting Eqs. (\ref{amplitude13})-(\ref{amplitude21}) into Eqs. (\ref{localized CP}) and (\ref{B}), we can directly obtain the $CP$ asymmetries and branching fractions of all the individual four-body decay channel $\bar{B}^0\rightarrow [K^-\pi^+]_{S/V}[\pi^+\pi^-]_{V/S} \rightarrow K^-\pi^+\pi^+\pi^-$ by integrating the phase space of Eq. (\ref{B}), respectively, both of which are summarized in Table \ref{CPviolationbr}. From this table, we can conclude that the range of these $CP$ asymmetries and branching fractions are about $[0.13, 29.4]\times10^{-2}$ and $[0.2, 38]\times10^{-6}$, respectively. Considering the contributions from all four-body decays listed in Table \ref{CPviolationbr}, we can obtain the localized integrated $CP$ asymmetries and branching fractions of the $\bar{B}^0\rightarrow K^-\pi^+\pi^+\pi^-$ decay by integrating the phase space. Our results are in the ranges $\mathcal{A_{CP}}(\bar{B}^0\rightarrow K^-\pi^+\pi^+\pi^-)=[-0.383,0.421]$ and $\mathcal{B}(\bar{B}^0\rightarrow K^-\pi^+\pi^+\pi^-)=[7.36,199.69]\times10^{-8}$ when the invariant masses of $K^-\pi^+$ and $\pi^-\pi^+$ are in the ranges $0.35<m_{K^-\pi^+}<2.04 \, \mathrm{GeV}$ and $0<m_{\pi^-\pi^+}<1.06\, \mathrm{GeV}$, where the $K\pi$ channel is dominated by the $\kappa$, $\bar{K}^*(892)^0$, $\bar{K}^*(1410)^0$, $\bar{K}^*_0(1430)$ and $\bar{K}^*(1680)^0$ resonances, and the $\pi\pi$ channel is dominated by the $\sigma$, $\rho^0(770)$ , $\omega(782)$ and $f_0(980)$ resonances, respectively, and the range of $\rho_A$ and $\phi_A$ are taken as $[0,0.5]$ and $[0,2\pi]$, respectively. Both of them are expected to be tested in the near future experiments.

\begin{table}[tb]
\renewcommand{\arraystretch}{1.2}
\centering
\caption{Direct $CP$ asymmetries (in units of $10^{-2}$) and branching fractions (in units of $10^{-6}$) of the four-body decays $\bar{B}^0\rightarrow [K^-\pi^+]_{S/V}[\pi^+\pi^-]_{V/S} \rightarrow K^-\pi^+\pi^+\pi^-$. The theoretical errors come from the uncertainties of the form factors, decay constants, Gegenbauer moments and divergence parameters.}
\begin{tabular*}{\textwidth}{@{\extracolsep{\fill}}ccc}
\hline
\hline
Decay mode                                                               &$CP$ asymmetries                                 & Branching fractions\\
\hline
$\bar{\kappa}$$\rho$$(\rightarrow K^-\pi^+\pi^+\pi^-)$                         &$-12.65\pm3.20$                               &$1.27\pm0.53$ \\
$\bar{\kappa}$$\omega$ $(\rightarrow K^-\pi^+\pi^+\pi^-)$                      &$15.19\pm4.96$                              &$4.12\pm0.66$\\
$\bar{K}^*(892)^0$$\sigma$ $(\rightarrow K^-\pi^+\pi^+\pi^-)$            &$27.82\pm8.60$                              &$0.13\pm0.07$\\
$\bar{K}^*(892)^0$$f_0(980)$ $(\rightarrow K^-\pi^+\pi^+\pi^-)$          &$-5.82\pm1.35$                              &$10.81\pm3.13$\\
$\bar{K}^*(1410)^0$$\sigma$ $(\rightarrow K^-\pi^+\pi^+\pi^-)$           &$0.27\pm0.39$                               &$22.58\pm6.35$\\
$\bar{K}^*(1410)^0$$f_0(980)$$(\rightarrow K^-\pi^+\pi^+\pi^-)$          &$-1.76\pm0.62$                               &$16.43\pm4.09$\\
$\bar{K}^*_0(1430)^0$$\rho$ $(\rightarrow K^-\pi^+\pi^+\pi^-)$           &$-5.89\pm1.81$                             &$1.96\pm0.39$\\
$\bar{K}^*_0(1430)^0$$\omega$ $(\rightarrow K^-\pi^+\pi^+\pi^-)$         &$13.48\pm4.20$                             &$2.72\pm0.94$\\
$\bar{K}^*(1680)^0$$\sigma$ $(\rightarrow K^-\pi^+\pi^+\pi^-)$           &$5.66\pm1.87$                             &$29.49\pm8.36$\\
$\bar{K}^*(1680)^0$$f_0(980)$$(\rightarrow K^-\pi^+\pi^+\pi^-)$          &$-2.98\pm0.82$                               &$20.34\pm4.54$\\
\hline
\hline
\end{tabular*}\label{CPviolationbr}
\end{table}

\begin{table}[tb]
\renewcommand{\arraystretch}{1.2}
\centering
\caption{The masses, widths and decay models of the intermediate resonances \cite{Tanabashi:2018oca}.}
\begin{tabular*}{\textwidth}{@{\extracolsep{\fill}}ccccc}
\hline
\hline
Resonance         &Mass ($\mathrm{MeV}$)   &Width ($\mathrm{MeV}$)               &$J^P$                       &Model\\
\hline
$\sigma$          & $475\pm75$               &$550\pm150$                                           &$0^+$                       &BUGG\\
$\rho$            &$775.26\pm0.25$           &$149.1\pm0.8$                                     &$1^-$                       &GS\\
$\omega$          &$782.65\pm0.12$           &$8.49\pm0.08$                                    &$1^-$                       &RBW\\
$f_0(980)$        &$990\pm20$                &$65\pm45$                                             &$0^+$                       &FLATT$\acute{\mathrm{E}}$\\
$\bar{\kappa}$          &$824\pm30$                &$478\pm50$                                          &$0^+$                       &RBW\\
$\bar{K}^*(892)^0$     &$895.5\pm0.20$            &$47.3\pm0.5$                                      &$1^-$                       &RBW\\
$\bar{K}^*(1410)^0$    &$1421\pm9$                &$236\pm18$                                          &$1^-$                       &RBW\\
$\bar{K}^*_0(1430)^0$    &$1425\pm50$               &$270\pm80$                                          &$0^+$                       &LASS\\
$\bar{K}^*(1680)^0$    &$1718\pm18$               &$322\pm110$                                          &$1^-$                       &RBW\\
\hline
\hline
\end{tabular*}\label{masses widths}
\end{table}

\section{SUMMARY}
In this work, we have revisited the four-body decay $\bar{B}^0\rightarrow K^-\pi^+\pi^-\pi^+$ in the framework of the two two-body decays. We consider more contributions from more different resonances. Meanwhile, we update the model when dealing with the dynamical function for the $\rho$ resonance.  The most important thing is that we have added the relevant calculations to further test the rationality of the two-quark model for scalar mesons in the two-body decay of the $\bar{B}^0$ meson. In this analysis, we first calculate the direct $CP$ violating asymmetries and branching fractions of the two-body decays $\bar{B}^0\rightarrow [K^-\pi^+]_{S/V}[\pi^+\pi^-]_{V/S}$ within the QCDF approach which are listed in  Table \ref{cpviolation1} and Table \ref{brachingfractions1}, respectively. From these two tables, we can see that our theoretical results are consistent with the available experimental data including the $CP$ asymmetries of the $\bar{B}^0\rightarrow \bar{K}^*(892)^0$$f_0(980)$ and $\bar{B}^0\rightarrow \bar{K}^*_0(1430)^0$$\omega$ decays and the branching fractions of the $\bar{B}^0\rightarrow \bar{K}^*(892)^0f_0(980)$, $\bar{B}^0\rightarrow\bar{K}^*_0(1430)^0\rho$ and $\bar{B}^0\rightarrow\bar{K}^*_0(1430)^0\omega$ decays. Because of different structures of the $\bar{K}^*_0(1430)^0$ meson, our predicted central values of the $CP$ asymmetries are larger than those in Ref. \cite{Cheng:2013fba} for the $\bar{B}^0\rightarrow \bar{K}^*_0(1430)^0 \rho$ and $\bar{B}^0\rightarrow \bar{K}^*_0(1430)^0\omega$ decays. It is found that the signs of the $CP$ asymmetries are negative for the $\bar{B}^0\rightarrow \bar{\kappa}\rho$, $\bar{B}^0\rightarrow \bar{K}^*(1410)^0 f_0(980)$ and $\bar{B}^0\rightarrow \bar{K}^*(1680)^0 f_0(980)$ decays and are positive for others decays. The magnitudes of branching fractions for our considered two-body decays $\bar{B}^0\rightarrow [K^-\pi^+]_{S/V}[\pi^+\pi^-]_{V/S}$ are of orders $10^{-7}\sim10^{-5}$. Then, under the assumption of the quasi-two-body decay mode, we regard $\bar{B}^0\rightarrow K^-\pi^+\pi^-\pi^+$ decay as happening through $\bar{B}^0\rightarrow [K^-\pi^+]_{S/V}[\pi^+\pi^-]_{V/S}\rightarrow K^-\pi^+\pi^-\pi^+$ and calculate the direct $CP$ asymmetries and branching fractions of all the individual four-body decay channel $\bar{B}^0\rightarrow [K^-\pi^+]_{S/V}[\pi^+\pi^-]_{V/S} \rightarrow K^-\pi^+\pi^+\pi^-$. The range of them are about $[0.13, 29.4]\times10^{-2}$ and $[0.2, 38]\times10^{-6}$, respectively. Finally, considering the contributions from all these decay channels decays, we obtain the localized integrated $CP$ asymmetries and the branching fraction of $\bar{B}^0\rightarrow K^-\pi^+\pi^-\pi^+$ when $0.35<m_{K^-\pi^+}<2.04 \, \mathrm{GeV}$ and $0<m_{\pi^-\pi^+}<1.06\, \mathrm{GeV}$, which are dominated by the $\bar{K}^*_0(700)^0$, $\bar{K}^*(892)^0$, $\bar{K}^*(1410)^0$, $\bar{K}^*_0(1430)$ and $\bar{K}^*(1680)^0$, and $f_0(500)$, $\rho^0(770)$ , $\omega(782)$ and $f_0(980)$ resonances, respectively. The predicted results are $\mathcal{A_{CP}}(\bar{B}^0\rightarrow K^-\pi^+\pi^+\pi^-)=[-0.383,0.421]$ and $\mathcal{B}(\bar{B}^0\rightarrow K^-\pi^+\pi^+\pi^-)=[7.36,199.69]\times10^{-8}$. In our analysis, the errors come from the uncertainties of the form factors, decay constants, Gegenbauer moments and divergence parameters. These theoretical predictions await the test in the future examinations with high precision. If our predictions are confirmed, the viewpoint that scalars have the $q\bar{q}$ composition may be supported. However, to exclude other possible structures, more investigations will be needed due to uncertainties from both theory and experiments.

\begin{appendix}
\section{FOUR-BODY DECAY AMPLITUDES}
Considering the related weak and strong decays, one can obtain the four-body decay amplitudes of the $\bar{B}^0\rightarrow [K^-\pi^+]_{S/V}[\pi^+\pi^-]_{V/S} \rightarrow K^-\pi^+\pi^+\pi^-$ channels as the following:
\begin{eqnarray}\label{amplitude13}
\begin{split}
\mathcal{M}(\bar{B}^0\rightarrow \bar{K}^{*0}_{0i}\rho\rightarrow K^-\pi^+\pi^+\pi^- )&=\frac{iG_Fg_{\bar{K}^{*0}_{0i}K\pi}g_{\rho\pi\pi}}{S_{\bar{K}^{*0}_{0i}}S_\rho }\bigg[(P\cdot N)+(L\cdot N)+\frac{1}{m_{\rho}^2}(L\cdot P+L^2)(L\cdot N)\bigg]\\
&\times\sum_{p=u,c}\lambda_p^{(s)}\bigg\{\bigg[\delta_{pu}\alpha_2(\bar{K}^{*0}_{0i}\rho)
+\frac{3}{2}\alpha_{3,EW}^p(\bar{K}^{*0}_{0i}\rho)\bigg]
f_\rho m_{\bar{B}^0}p_cF_1^{\bar{B}^0 \bar{K}^{*0}_{0i}}(m_{\rho}^2)\\
&+\bigg[\alpha_4^p(\rho\bar{K}^{*0}_{0i})-\frac{1}{2}\alpha_{4,EW}^p(\rho\bar{K}^{*0}_{0i})\bigg]
\bar{f}_{\bar{K}^{*0}_{0i}}m_{\bar{B}^0}p_cA_0^{\bar{B}^0\rho}(m_{\bar{K}^{*0}_{0i}}^2)\\
&+\bigg[\frac{1}{2}b_3^p(\rho\bar{K}^{*0}_{0i})-\frac{1}{4}b_{3,EW}^p(\rho\bar{K}^{*0}_{0i})\bigg]
\frac{f_{\bar{B}^0}f_\rho \bar{f}_{\bar{K}^{*0}_{0i}}m_{\bar{B}^0}p_c}{m_\rho}\bigg\},\\
\end{split}
\end{eqnarray}

\begin{eqnarray}\label{amplitude142}
\begin{split}
\mathcal{M}(\bar{B}^0\rightarrow \bar{K}^{*0}_{0i}\omega\rightarrow K^-\pi^+\pi^+\pi^- )&=\frac{iG_Fg_{\bar{K}^{*0}_{0i}K\pi}g_{\omega\pi\pi}}{S_{\bar{K}^{*0}_{0i}}S_\omega}\bigg[(P\cdot N)+(L\cdot N)+\frac{1}{m_{\omega}^2}(L\cdot P+L^2)(L\cdot N)\bigg]\\
&\times\sum_{p=u,c}\lambda_p^{(s)}\bigg\{\bigg[\delta_{pu}\alpha_2(\bar{K}^{*0}_{0i}\omega)+2\alpha_3^p(\bar{K}^{*0}_{0i}\omega)
+\frac{1}{2}\alpha_{3,EW}^p(\bar{K}^{*0}_{0i}\omega)\bigg]\\
&\times f_\omega m_{\bar{B}^0}p_cF_1^{\bar{B}^0 \bar{K}^{*0}_{0i}}(m_{\omega}^2)
+\bigg[\frac{1}{2}\alpha_{4,EW}^p(\omega\bar{K}^{*0}_{0i})-\alpha_4^p(\omega\bar{K}^{*0}_{0i})\bigg]\bar{f}_{\bar{K}^{*0}_{0i}}m_{\bar{B}^0}p_c\\
&\times A_0^{\bar{B}^0\omega}(m_{\bar{K}^{*0}_{0i}}^2)
+\bigg[\frac{1}{4}b_{3,EW}^p(\omega\bar{K}^{*0}_{0i})-\frac{1}{2}b_3^p(\omega\bar{K}^{*0}_{0i})\bigg]
\frac{f_{\bar{B}^0}f_\rho \bar{f}_{\bar{K}^{*0}_{0i}}m_{\bar{B}^0}p_c}{m_\omega}\bigg\},\\
\end{split}
\end{eqnarray}
and
\begin{eqnarray}\label{amplitude21}
\begin{split}
\mathcal{M}(\bar{B}^0\rightarrow \bar{K}^{*0}_if_{0j}\rightarrow K^-\pi^+\pi^+\pi^-)&=-\frac{iG_Fg_{\bar{K}^{*0}_iK\pi}g_{f_{0j}\pi\pi}}{S_{\bar{K}^{*0}_i}S_{f_{0j}}}
\bigg[-(P\cdot Q)-(L\cdot Q)+\frac{1}{{m_{\bar{K}^{*0}_i}}^2}(P^2+P\cdot L)(P\cdot Q)]\bigg]\\
&\sum_{p=u,c}\lambda_p^{(s)}\bigg\{\bigg[\delta_{pu}\alpha_2(\bar{K}^{*0}_if_{0j})+2\alpha_3^p(\bar{K}^{*0}_if_{0j})
+\frac{1}{2}\alpha_{3,EW}^p(\bar{K}^{*0}_if_{0j})\bigg]\\
&\times \bar{f}_{f_{0j}^n} m_{\bar{B}^0}p_cA_0^{\bar{B}^0 \bar{K}^{*0}_i}(m_{f_{0j}}^2)+\bigg[\sqrt{2}\alpha_3^p(\bar{K}^{*0}_if_{0j})+\sqrt{2}\alpha_4^p(\bar{K}^{*0}_if_{0j})\\
&-\frac{1}{\sqrt{2}}\alpha_{3,EW}^p(\bar{K}^{*0}_i\sigma)-\frac{1}{\sqrt{2}}\alpha_{4,EW}^p(\bar{K}^{*0}_i\sigma)\bigg]\bar{f}_{\sigma^s} m_{\bar{B}^0}p_cA_0^{\bar{B}^0\bar{K}^{*0}_i}(m_{f_{0j}}^2)\\
&+\bigg[\frac{1}{2}\alpha_{4,EW}^p(f_{0j}\bar{K}^{*0}_i)-\alpha_4^p(f_{0j}\bar{K}^{*0}_i)\bigg]f_{\bar{K}^{*0}_i}m_{\bar{B}^0}p_c
F_1^{\bar{B}^0f_{0j}}(m_{\bar{K}^{*0}_i}^2)\\
&+\bigg[\frac{1}{\sqrt{2}}b_3^p(\bar{K}^{*0}_if_{0j})-\frac{1}{2\sqrt{2}}b_{3,EW}^p(\bar{K}^{*0}_if_{0j})\bigg]
\frac{f_{\bar{B}^0}f_{\bar{K}^{*0}_i}\bar{f}_{f_{0j}}^s m_{\bar{B}^0}p_c}{m_{\bar{K}^{*0}_i}}\\
&+\bigg[\frac{1}{2}b_3^p(f_{0j}\bar{K}^{*0}_i)-\frac{1}{4}b_{3,EW}^p(f_{0j}\bar{K}^{*0}_i)\bigg]\frac{f_{\bar{B}^0}f_{\bar{K}^{*0}_i}\bar{f}_{f_{0j}}^nm_{\bar{B}^0}p_c}{m_{\bar{K}^{*0}_i}}\bigg\}.\\
\end{split}
\end{eqnarray}

\section{DYNAMICAL FUNCTIONS FOR THE CORRESPONDING RESONANCES}
\subsection{BUGG MODEL}
For the $\sigma$ resonance, we adopt the Bugg model \cite{Bugg:2006gc} for parameterization
\begin{equation}\label{sigma1}
\begin{split}
 T_R(m_{\pi\pi})=1/[M^2-s_{\pi\pi}-g_1^2(s_{\pi\pi})\frac{s_{\pi\pi}-s_A}{M^2-s_A}z(s_{\pi\pi})-iM\Gamma_{\mathrm{tot}}(s_{\pi\pi})],
 \end{split}
 \end{equation}
 where $z(s_{\pi\pi})=j_1(s_{\pi\pi})-j_1(M^2)$ with $j_1(s_{\pi\pi})=\frac{1}{\pi}[2+\rho_1\ln(\frac{1-\rho_1}{1+\rho_1})]$, $\Gamma_{\mathrm{tot}}(s_{\pi\pi})=\sum\limits _{i=1}^4 \Gamma_i(s_{\pi\pi})$ with
\begin{equation}\label{sigma2}
\begin{split}
M\Gamma_1(s_{\pi\pi})&=g_1^2(s_{\pi\pi})\frac{s_{\pi\pi}-s_A}{M^2-s_A}\rho_1(s_{\pi\pi}),\\
M\Gamma_2(s_{\pi\pi})&=0.6g_1^2(s_{\pi\pi})(s_{\pi\pi}/M^2)\mathrm{exp}(-\alpha|s_{\pi\pi}-4m_K^2|)\rho_2(s_{\pi\pi}),\\
M\Gamma_3(s_{\pi\pi})&=0.2g_1^2(s_{\pi\pi})(s_{\pi\pi}/M^2)\mathrm{exp}(-\alpha|s_{\pi\pi}-4m_\eta^2|)\rho_3(s_{\pi\pi}),\\
M\Gamma_4(s_{\pi\pi})&=Mg_4\rho_{4\pi}(s_{\pi\pi})/\rho_{4\pi}(M^2),\\
\end{split}
 \end{equation}
and
\begin{equation}\label{sigma3}
\begin{split}
g_1^2(s_{\pi\pi})&=M(b_1+b_2s)\mathrm{exp}[-(s_{\pi\pi}-M^2)/A],\\
\rho_{4\pi}(s_{\pi\pi})&=1.0/[1+\mathrm{exp}(7.082-2.845s_{\pi\pi})].\\
 \end{split}
 \end{equation}

For the parameters in Eqs. (\ref{sigma1}, \ref{sigma2}), they are fixed as $M=0.953\,\mathrm{GeV}$, $s_A=0.14m_\pi^2$, $b_1=1.302 \,\mathrm{GeV}^2$, $b^2=0.340$,$A=2.426\, \mathrm{GeV}^2$ and $g_{4\pi}=0.011\, \mathrm{GeV}$ which are given in the fourth column of Table I in Ref. \cite{Bugg:2006gc}. The parameters $\rho_{1,2,3}$ are the phase-space factors of the decay channels $\pi\pi$, $KK$ and $\eta\eta$, respectively, and have been defined as \cite{Bugg:2006gc}
\begin{equation}\label{rho}
\rho_i(s_{\pi\pi})=\sqrt{1-4\frac{m_i^2}{s_{\pi\pi}}},
\end{equation}
with $m_1=m_\pi$, $m_2=m_K$ and $m_3=m_\eta$.

\subsection{THE GOUNARIS-SAKURAI FUNCTION}
For the $\rho^0(770)$ resonance, an analytic dispersive term is included to ensure unitarity far from the pole mass, known as the Gounaris-Sakurai model. It takes the
form \cite{Gounaris:1968mw}

\begin{equation} \label{TR43}
\begin{split}
T_R(m_{\pi\pi})=\frac{1+D\Gamma_0/m_0}{m_0^2-s_{\pi\pi}+f(m_{\pi\pi})-im_0\Gamma(m_{\pi\pi})},\\
 \end{split}
 \end{equation}
where $\Gamma_0$ and $m_0$ are the natural width and the Breit-Wigner mass of the $\rho^0(770)$ meson, respectively, The concrete form of $f(m_{\pi\pi})$ is
\begin{equation} \label{TR44}
\begin{split}
f(m_{\pi\pi})=\Gamma_0\frac{m_0^2}{q_0^3}\left[q^2\left[h(m_{\pi\pi})-h(m_0)\right]+(m_0^2-m_{\pi\pi}^2)q^2_0\frac{\mathrm{d}h}{\mathrm{d}m_{\pi\pi}^2}\bigg|_{m_0}\right],\\
 \end{split}
 \end{equation}
where $q_0$ is the value of $q=|\vec{q}|$ when the invariant mass, $m_{\pi\pi}$, is equal to the mass of the $\rho^0(770)$ resonance, with
\begin{equation} \label{TR45}
\begin{split}
h(m_{\pi\pi})=\frac{2}{\pi}\frac{q}{m_{\pi\pi}}\log\bigg(\frac{m_{\pi\pi}+2q}{2m_\pi}\bigg),\\
 \end{split}
 \end{equation}

\begin{equation} \label{TR46}
\begin{split}
\frac{\mathrm{d}h}{\mathrm{d}m_{\pi\pi}^2}\bigg|_{m_0}=h(m_0)\left[(8q_0^2)^{-1}-(2m_0^2)^{-1}\right]+(2\pi m_0^2)^{-1}.\\
 \end{split}
 \end{equation}

 The constant parameter $D$ is given by

 \begin{equation} \label{TR47}
\begin{split}
D=\frac{3}{\pi}\frac{m_\pi^2}{q_0^2}\log\bigg(\frac{m_0+2q_0}{2m_\pi}\bigg)+\frac{m_0}{2\pi q_0}-\frac{m_\pi^2 m_0}{\pi q_0^3}.\\
 \end{split}
 \end{equation}

\subsection{FLATT$\acute{\mathrm{E}}$ MODEL}
As suggested by D. V. Bugg \cite{Bugg:2008ig}, the Flatt$\acute{\mathrm{e}}$ model \cite{Flatte:1976xv} for $f_0(980)$ is slightly modified and is parametrized as
\begin{equation}\label{rho}
T_R(m_{\pi\pi})=\frac{1}{m_R^2-s_{\pi\pi}-im_R(g_{\pi\pi}\rho_{\pi\pi}+g_{KK}F_{KK}^2\rho_{KK})},
\end{equation}
where $m_R$ is the $f_0(980)$ pole mass, the parameters $g_{\pi\pi}$ and $g_{KK}$ are the $f_0(980)$ coupling constants with respect to the $\pi^+\pi^-$ and $K^+K^-$ final states, respectively, and the phase-space $\rho$ factors are given by Lorentz-invariant phase spaces as

\begin{equation}\label{f01}
\begin{split}
\rho_{\pi\pi}&=\frac{2}{3}\sqrt{1-\frac{4m_{\pi^\pm}^2}{s_{\pi\pi}}}+\frac{1}{3}\sqrt{1-\frac{4m_{\pi^0}^2}{s_{\pi\pi}}},\\
\rho_{KK}&=\frac{1}{2}\sqrt{1-\frac{4m_{K^\pm}^2}{s_{\pi\pi}}}+\frac{1}{2}\sqrt{1-\frac{4m_{K^0}^2}{s_{\pi\pi}}}.\\
 \end{split}
 \end{equation}
In Eq. (\ref{rho}), compared to the normal Flatt$\acute{\mathrm{e}}$ function, a form factor $F_{KK}=\mathrm{exp}(-\alpha k^2)$ is introduced above the $KK$ threshold and serves to reduce the $\rho_{KK}$ factor as $s_{\pi\pi}$ increases, where $k$ is momentum of each kaon in the $KK$ rest frame, and $\alpha=(2.0\pm0.25)\,\mathrm{GeV}^{-2}$ \cite{Bugg:2008ig}. This parametrization slightly decreases the $f_0(980)$ width above the $KK$ threshold. The parameter $\alpha$ is fixed to be $2.0 \,\mathrm{GeV}^{-2}$, which is not very sensitive to the fit.

\subsection{LASS MODEL}
The S-wave $K^+\pi^-$ resonance at low mass is modeled using a modified LASS lineshape \cite{Aston:1987ir,Aubert:2005ce,Aaij:2018bla}, which has been widely used in experimental analyses:
\begin{equation}\label{K1430 1}
\begin{split}
T(m_{K\pi})&=\frac{m_{K\pi}}{|\vec{q}|\cot\delta_B-i|\vec{q}|}+e^{2i\delta_B}\frac{m_0\Gamma_0\frac{m_0}{|q_0|}}{m_0^2-s_{K\pi}^2-im_0\Gamma_0\frac{|\vec{q}|}{m_{K\pi}}\frac{m_0}{|q_0|}},\\
 \end{split}
 \end{equation}
with
\begin{equation}\label{K1430 2}
\begin{split}
\cot\delta_B&=\frac{1}{a|\vec{q}|}+\frac{1}{2}r|\vec{q}|,\\
 \end{split}
 \end{equation}
where the first term is an empirical term from the elastic kaon-pion scattering and the second term is the resonant contribution with a phase factor to retain unitarity. Here $m_0$ and $\Gamma_0$ are the pole mass and width of the $K_0^*(1430)$ state, respectively, $|\vec{q}|$ is the momentum vector of the resonance decay product measured in the resonance rest frame, and $|\vec{q_0}|$ is the value of $|\vec{q}|$ when $m_{K\pi}=m_{K_0^*(1430)}$. In Eq. (\ref{K1430 2}), the parameters $a=(3.1\pm1.0)\,\mathrm{GeV}^{-1}$ and $r=(7.0\pm2.3)\,\mathrm{GeV}^{-1}$ are the scattering length and effective range \cite{Aaij:2018bla}, respectively, which are universal in application for the description of different processes involving a kaon-pion pair.

\subsection{RELATIVISTIC BREIT-WIGNER}
We adopt the relativistic Breit-Wigner function to describe the distributions of the $\bar{K}^*_0(700)^0$, $\bar{K}^*(892)^0$, $\bar{K}^*(1410)^0$ and $\bar{K}^*(1680)^0$ resonances \cite{Chen:2020sog},
\begin{equation} \label{TR}
\begin{split}
T_R(m_{K\pi})=\frac{1}{M_R^2-s_{K\pi}-iM_R\Gamma_{K\pi}}  \quad\quad\quad(R=\bar{\kappa},\bar{K}^*),
 \end{split}
 \end{equation}
 with
 \begin{equation} \label{TRr}
\begin{split}
\Gamma_{K\pi}=\Gamma_0^R\bigg(\frac{p_{K\pi}}{p_R}\bigg)^{2J+1}(\frac{M_R}{m_{K\pi}})F^2_R,
 \end{split}
 \end{equation}
where $m_{K\pi}$ is the invariant mass of the $K\pi$ pair, $s_{K\pi}=m_{K\pi}^2$, $p_{K\pi}(p_R)$ is the momentum of either daughter in the $K\pi$ (or $R$) rest frame, and $M_R$ and $\Gamma_0^R$ are the nominal mass and width, respectively, $F_R$ is the Blatt-Weisskopf centrifugal barrier factor \cite{J Blatt}, which are listed in Table \ref{FR} and depend on a single parameter $R_r$ representing the meson radius, for which one can adopt $R_r=1.5\mathrm{GeV}^{-1}$ \cite{Kopp:2000gv}.
\begin{table}[hptb]
\centering
\caption{
Summary of the Blatt-Weisskopf penetration form factors.}
\begin{tabular}{p{3.7cm}<{\centering} p{7.7cm}<{\centering}}
\hline
\hline
 Spin&  $F_R$   \\
\hline
0&      1   \\
1&    $\frac{\sqrt{1+(R_r p_R)^2}}{\sqrt{1+(R_r p_{AB})^2}}$\\
\hline
\hline
\end{tabular}\label{FR}
\end{table}

\section{$f_0(500)-f_0(980)$ MIXING}
 Analogous to the $\eta-\eta'$ mixing, the scalar $f_0(500)-f_0(980)$
mixing can also be parameterized by a $2 \times 2$ rotation matrix
with a single angle $\varphi_m$ in the quark-flavor basis, namely,

\begin{equation}
\left(
\begin{array}{cc}
f_0(980)\\
 f_0(500)\\
\end{array}
\right)=
\left(
\begin{array}{cc}
 \cos\varphi_m& \sin\varphi_m \\
 -\sin\varphi_m&  \cos\varphi_m
\end{array}
\right )
\left(
\begin{array}{cc}
 f_s\\
f_q\\
\end{array}
\right )
\end{equation}
with the quark-flavor states $f_s\equiv s\bar{s}$ and $f_q\equiv \frac{u\bar{u}+d\bar{d}}{\sqrt{2}}$. Various mixing angle $\varphi_m$ measurements have been derived and summarized in the literature with a wide range of values \cite{Fleischer:2011au,Cheng:2013fba}. However, it is worth pointing out that, based on the recent measurement and the accompanied discussion performed by the LHCb Collaboration \cite{Aaij:2013zpt}, the upper limit $|\varphi_m|<31^0$ has been set for the first time in the $B$ meson decays with a two-quark structure description of $f_0(500)$ and $f_0(980)$. In our calculation, we adopt $|\varphi_m|=17^0$ \cite{Cheng:2013fba}.

\section{THEORETICAL INPUT PARAMETERS}
The predictions obtained in the QCDF  approach depend on many input parameters. The values of the Wolfenstein parameters are taken from Ref. \cite{Agashe:2014kda}: $\bar{\rho}=0.117\pm0.021$, $\bar{\eta}=0.353\pm0.013$.

The effective Wilson coefficients used in our calculations are taken from Ref. \cite{Qi:2018syl}:
\begin{equation}\label{C}
\begin{split}
&C'_1=-0.3125, \quad C'_2=-1.1502, \\
&C'_3=2.120\times10^{-2}+5.174\times10^{-3}i,\quad C'_4=-4.869\times10^{-2}-1.552\times10^{-2}i, \\
&C'_5=1.420\times10^{-2}+5.174\times10^{-3}i,\quad C'_6=-5.792\times10^{-2}-1.552\times10^{-2}i, \\
&C'_7=-8.340\times10^{-5}-9.938\times10^{-5}i,\quad C'_8=3.839\times10^{-4}, \\
&C'_9=-1.017\times10^{-2}-9.938\times10^{-5}i,\quad C'_{10}=1.959\times10^{-3}. \\
\end{split}
\end{equation}

For the masses used in $\bar{B}^0$ decays, we use the following values except those listed in Table \ref{masses widths} (in $\mathrm{GeV}$) \cite{Agashe:2014kda}:
\begin{equation}
\begin{split}
m_u&=m_d=0.0035,\quad m_s=0.119, \quad m_b=4.2,\\
m_{\pi^\pm}&=0.14,\quad m_{K^-}=0.494,\quad m_{\bar{B}^0}=5.28,\\
\end{split}
\end{equation}
while for the widths we shall use (in units of $\mathrm{GeV}$) \cite{Agashe:2014kda}
\begin{equation}
\begin{split}
\Gamma_{\rho\rightarrow\pi\pi}&=0.149,\quad\Gamma_{\omega\rightarrow\pi\pi}=0.00013,\quad\Gamma_{\sigma\rightarrow\pi\pi}=0.3,\quad \Gamma_{f_0(980)\rightarrow \pi\pi}=0.33,\\
\Gamma_{\bar{K}^*(892)^0\rightarrow K\pi}&=0.0487,\quad\Gamma_{\bar{K}^*(1410)^0\rightarrow K\pi}=0.015,\quad \Gamma_{\bar{K}^*(1680)^0\rightarrow K\pi}=0.10,\quad \Gamma_{K^*_0(1430)\rightarrow K\pi}=0.251.\\
\end{split}
\end{equation}

The strong coupling constants are determined from the measured partial widths through the relations \cite{Cheng:2013dua,Dedonder:2014xpa}
\begin{equation}\label{gSV}
\begin{split}
g_{S M_1M_2}=\sqrt{\frac{8\pi m_S^2}{p_c(S)}\Gamma_{S\rightarrow M_1M_2}},\\
g_{V M_1M_2}=\sqrt{\frac{6\pi m_V^2}{p_c(V)^3}\Gamma_{V\rightarrow M_1M_2}},\\
\end{split}
\end{equation}
where $p_c(S,V)$ are the magnitudes of the three momenta of the final state mesons in the rest frame of $S$ and $V$ mesons, respectively.

The following relevant decay constants (in $\mathrm{GeV}$) are used \cite{Cheng:2010yd,Cheng:2005nb,Cheng:2005ye}:
\begin{equation}
\begin{split}
f_{\pi^\pm}&=0.131,\quad f_{\bar{B}^0}=0.21\pm0.02, \quad f_{K^-}=0.156\pm0.007,  \\
\bar{f}^s_{\sigma}&=-0.21\pm0.093,\quad \bar{f}_{\sigma}^u=0.4829\pm0.076,\quad \bar{f}_{\bar{\kappa}}=0.34\pm0.02,\\
f_{\rho}&=0.216\pm0.003,\quad f_{\rho}^\perp=0.165\pm0.009,\\
f_{\omega}&=0.187\pm0.005,\quad f_{\omega}^\perp=0.151\pm0.009,\\
f_{\bar{K}^*(892)^0}&=0.22\pm0.005,\quad f_{\bar{K}^*(892)^0}^\perp=0.185\pm0.010,\quad \bar{f}_{\bar{K}^*_0(1430)^0}=-0.300\pm0.030. \\
\bar{f}^s_{f_0(980)}&=0.325\pm0.016,\quad \bar{f}_{f_0(980)}^u=0.1013\pm0.005.\\
\end{split}
\end{equation}

As for the form factors, we use \cite{Cheng:2010yd,Cheng:2005nb,Deandrea:2000ce}:
\begin{equation}
\begin{split}
F_0^{\bar{B}^0\rightarrow K}(0)&=0.35\pm0.04,\quad F_0^{\bar{B}^0\rightarrow \sigma}(0)=0.45\pm0.15,\quad A_0^{\bar{B}^0\rightarrow \rho}(0)=0.303\pm0.029,\\
A_0^{\bar{B}^0\rightarrow \bar{K}^*(892)^0}(0)&=0.374\pm0.034, \quad F_0^{\bar{B}^0\rightarrow \pi}(0)=0.25\pm0.03, \quad F_0^{\bar{B}^0\rightarrow \bar{K}^*_0(1430)^0}(0)=0.21.\\
\end{split}
\end{equation}

The values of Gegenbauer moments at $\mu=1 \mathrm{GeV}$ are taken from \cite{Cheng:2010yd,Cheng:2005nb,Cheng:2005ye},
\begin{equation}
\begin{split}
\alpha_1^\rho&=0,\quad \alpha_2^\rho=0.15\pm0.07, \quad \alpha_{1,\perp}^\rho=0,\quad \alpha_{2,\perp}^\rho=0.14\pm0.06, \\
\alpha_1^\omega&=0,\quad \alpha_2^\omega=0.15\pm0.07, \quad \alpha_{1,\perp}^\omega=0,\quad \alpha_{2,\perp}^\omega=0.14\pm0.06, \\
\alpha_1^{\bar{K}^*(892)^0}&=0.03\pm0.02,\quad \alpha_{1,\perp}^{\bar{K}^*(892)^0}=0.04\pm0.03,\\
\alpha_2^{\bar{K}^*(892)^0}&=0.11\pm0.09,\quad \alpha_{2,\perp}^{\bar{K}^*(892)^0}=0.10\pm0.08,\\
B_{1,\sigma}^u&=-0.42\pm0.074,\quad B_{3,\sigma}^u=-0.58\pm0.23,\\
B_{1,\sigma}^s&=-0.35\pm0.061,\quad B_{3,\sigma}^s=-0.43\pm0.18,\\
B_{1,f_0(980)}^u&=-0.92\pm0.08,\quad B_{3,f_0(980)}^u=-0.74\pm0.064,\\
B_{1,f_0(980)}^s&=-1\pm0.05,\quad B_{3,f_0(980)}^s=-0.8\pm0.04,\\
B_{1,\bar{\kappa}}&=-0.92\pm0.11,\quad B_{3,\bar{\kappa}}=0.15\pm0.09,\\
B_{1,\bar{K}^*_0(1430)^0}&=0.58\pm0.07,\quad B_{3,\bar{K}^*_0(1430)^0}=-1.20\pm0.08.\\
\end{split}
\end{equation}

\end{appendix}
\acknowledgments
This work was supported by National Natural Science Foundation of China (Projects Nos. 11575023, 11775024, 11947001, 11605150 and 11805153), Achievements of the Basic Scientific Research Business Foundation Project of Universities in Zhejiang Province and Ningbo Natural Science Foundation (No. 2019A610067).

\end{document}